\documentclass[aps,preprint]{revtex4}%
\usepackage{amsfonts}
\usepackage{amsmath}
\usepackage{amssymb}
\usepackage{graphicx}%

\begin{document}
\preprint{ }

\title{Dynamics of algebras in quantum unstable systems}

\author{Marcelo Losada}

\affiliation{Universidad de Buenos Aires - Consejo Nacional de Investigaciones
Cient\'ificas y T\'ecnicas, Ciudad de Buenos Aires, Argentina}

\author{Sebastian Fortin}
\affiliation{CONICET - Universidad de Buenos Aires, Ciudad de Buenos Aires, Argentina}

\author{Manuel Gadella}
\affiliation{Departamento de F\'isica Te\'orica, At\'omica  y \'Optica and IMUVA, Universidad de Valladolid \\ Paseo Bel\'en 7, 47011 Valladolid, Spain}

\author{Federico Holik\footnote{All authors contributed equally to this manuscript.}}
\affiliation{Instituto de F\'isica La Plata - Consejo Nacional de Investigaciones
Cient\'ificas y T\'ecnicas, La Plata, Buenos Aires, Argentina}

\begin{abstract}
We introduce a dynamical evolution operator for dealing with unstable
physical process, such as scattering resonances, photon emission,
decoherence and particle decay. With that aim, we use the formalism of
rigged Hilbert space and represent the time evolution of quantum
observables in the Heisenberg picture, in such a way that time evolution is
non-unitary. This allows to describe observables that are initially
non-commutative, but become commutative after time evolution. In other
words, a non-abelian algebra of relevant observables becomes abelian when times goes to infinity. We finally present some relevant examples.
\end{abstract}
\date{February 2013}
\maketitle

\section{Introduction}

In previous papers \cite{Fortin2014,Fortin2016,Losada2017}, we considered
the quantum-to-classical transition from the point of view of the algebra of
quantum observables. If a quantum system undergoes a physical process such
that its behavior becomes classical, then its algebra of observables should
undergo a transition from a non-Abelian algebra to an Abelian one. In order to
describe this kind of time evolutions, we have proposed to use the
Heisenberg picture, so we can consider the time evolution of the whole
algebra of observables. It is important to remark that, in the standard
formalism of quantum mechanics, a closed system always evolves unitarily.
So, even in the Heisenberg picture, if two observables are
incompatible at one time, they will remain incompatible for every time.
Therefore, with the aim of describing the quantum-to-classical transition of
the algebra of observables, it is necessary to go beyond unitary time
evolutions.

In this paper, we continue with this approach by studying more general
models. We introduce a dynamical evolution operator for dealing with
unstable physical process (such as scattering resonances, photon emission,
decoherence, relaxations and particle decay). In order to study the time
evolution of their algebras of observables, we use the formalism of
rigged Hilbert space (RHS), which is a natural choice for describing
these kind of systems. The RGS description of quantum mechanics is an
alternative formalism to that of von Neumann. It has several applications,
particularly in particle physics and in the study of scattering processes.
It also provides a rigorous description of eigenstates of the position and
momentum operators, in fact, it serves as a rigorous mathematical basis for
the Dirac formulation of quantum mechanics \cite{BOH,ROB,ANT,MELSH,GG,GG1}.

As mentioned above, the use of the Heisenberg picture allows to study the classical limit from a different point of view. We show that an initially
non-abelian algebra of relevant observables, becomes an abelian one when
times goes to infinity. We refer to this non-abelian/abelian transition as
\emph{commutation process}. The study of this process focuses on
the dynamics of the algebras of observables. In this work we provide an
explicit representation of the time evolution operator for an extensive
family of models described by the RHS formalism. We show that, under certain
conditions, a commutation process (of the form described in \cite%
{Fortin2014,Fortin2016,Losada2017}) is obtained for them. This phenomenon
could be of interest for the study of quantum scattering resonances. It
consists in a scattering process in which the scattered particle ends up in
a quasi-stationary state. As a result of our work, it turns out that the use
of non-Hermitian Hamiltonians of the form $H+\lambda V$, introduces a
natural ground for the study of the commutation process of algebras.

The paper is organized as follows. In Section \ref{SECII}, we introduce the
problem of the dynamical evolution of algebras and the logical
quantum-to-classical transition. We illustrate our ideas by discussing a
simple case: quantum operations and the quantum damping channel.
Next, in Section \ref{s:RHS}, we discuss the fundamental aspects of the RHS
formalism. In Section \ref{s:TimeEvolutionInRHS}, we introduce a time
evolution operator for observables in the RHS formalism. This allows us to
describe the commutation process for a family of models of unstable
systems in Section \ref{s:Commutativization}.

\section{Logical quantum-to-classical transition}

\label{SECII}

The sets of properties of classical and quantum systems have a logical
structure, given by their orthocomplemented lattice structure \cite{Jauch}
(see also \cite{Holik2010,Holik2012,Holik2013} for a recent discussion on
the subject). The propositional approach to quantum systems has been used in
diverse areas of the foundations of quantum mechanics, as for example, in
the study of quantum histories \cite{Omn 94, Gri 2014, LVL2013,
LL2014, LL2014b, LVL2016, Losada2018}. Due to this structure, logical operations and
logical relations between properties can be defined, such as disjunction ($%
\vee$), conjunction ($\wedge$), negation ($\neg$) and implication ($\leq$).
All orthocomplemented lattices satisfy certain relations, called \textit{%
distributive inequalities} \cite{Kalmbach}:
\begin{equation}
a\wedge(b\vee c)\geq (a\wedge b)\vee(a\wedge c), ~~~~~~ a\vee(b\wedge c)\geq
(a\vee b)\wedge(a\vee c),
\end{equation}
where $a$, $b$, and $c$ are arbitrary properties of the system. When the
equalities hold, the lattice is called \emph{distributive}. An
orthocomplemented and distributive lattice is called a \emph{Boolean lattice}%
. The distributive property is an essential feature which differentiates
classical and quantum lattices.

In classical mechanics a physical system is represented by a phase space and
the properties of the system are represented by measurable subsets of its
phase space. Therefore, the logical structure of classical systems is given
by the algebraic structure of sets \cite{Jauch}. The resulting lattice is
not only an orthocomplemented lattice, but also a distributive one, i.e., it
is a Boolean lattice. This logical structure is naturally related with \emph{%
classical logic}.

The quantum case is very different. In quantum mechanics a physical system
is represented by a Hilbert space, observables are represented by
self-adjoint operators on the Hilbert space and physical properties are
represented by orthogonal projectors \cite{von Neumann}. The logical
structure of quantum systems is the algebraic structure of orthogonal
projectors, and it is known as \textit{quantum logic} \cite{Birkhoff,Cohen}.

Unlike classical logic, quantum logic is a non-distributive
orthocomplemented lattice. While in the classic lattice, all properties
satisfy the distributive equalities, in the quantum lattice, only
distributive inequalities hold in general \cite{Jauch,Kalmbach}. However,
for some subsets of quantum properties the equalities hold. When a subset of
properties satisfies the distributive equalities, they are called \textit{%
compatible properties}. It can be proved that properties associated with
different observables are compatible if the observables commute. If, on the
contrary, two observables do not commute, some of the properties associated
with them are not compatible \cite{Jauch,Cohen}. Therefore, by extension,
commuting observables are called \textit{compatible observables}.

The differences between classical and quantum logic are of fundamental
importance for describing the quantum-to-classical transition. If a quantum
system undergoes a physical process and as a consequence of this its
behaviour becomes classic, then the logical structure of its properties
should undergo a transition from quantum logic to classical logic, i.e. its
lattice structure should become distributive. In order to give an adequate
description of the logical structure transition, we have proposed to
describe the classical limit in terms of the Heisenberg picture \cite%
{Fortin2014,Fortin2016,Losada2017}. This perspective allows to consider the
time evolution of the whole lattice of properties, and to study the
transition from classical to quantum logic.

It is important to remark that, when governed by the Schr\"odinger equation,
the time evolution of a closed system is always unitary. Even in the
Heisenberg picture, if two observables are incompatible at one time, they
will remain incompatible at any time \cite{Losada2017}. Therefore, for
describing the logical quantum-to-classical transition, it is necessary to
consider more general time evolutions.

In order to describe adequately the logical quantum-to-classical transition,
let us consider a quantum system with a general time evolution, and a
time-dependent set of relevant observables, $\mathcal{O}(t)= \{ \hat{O}_i(t)
\}_{i\in I}$ ($I$ an arbitrary set of indexes). Each set $\mathcal{O}(t)$
generates an algebra of observables $\mathcal{V}(t)$, and each algebra has
associated an orthocomplemented lattice $\mathcal{L}_{\mathcal{V}(t)}$. We
assume that initially some observables are incompatible, i.e., there are $%
i,j\in I$ such that $\left[\hat{O}_i(0), \hat{O}_j(0) \right]\neq 0$.
Therefore, the lattice $\mathcal{L}_{\mathcal{V}(0)}$ is a non-distributive
lattice.

For quantum systems with only one characteristic time $t_c$, the
quantum-to-classical transition is given by the following process:
\begin{equation}
\left[ \hat{O}_i(0), \hat{O}_j(0) \right]\neq 0 \longrightarrow \left[ \hat{O%
}_i(t_c), \hat{O}_j(t_c) \right]= 0, ~~~~ \forall i,j.
\end{equation}
The logical classical limit is expressed by the fact that, while $\mathcal{L}%
_{\mathcal{V}(0)}$ is a non-distributive lattice, $\mathcal{L}_{\mathcal{V}%
(t_c)}$ is a Boolean one, i.e., it is a classical logic. In this way, we
obtain an adequate description of the logical evolution of a quantum system.

In order to illustrate the general idea of the logical classical limit, we
are going to show the logical transition of a physical with a quantum
evolution given by a quantum channel.

\subsection{A simple case: quantum operations}

\label{SECIII}

We consider a time evolution given by a quantum operation, and we define the
corresponding Heisenberg representation. Once we have defined the quantum
operations on the space of quantum observables, we study the logical
quantum-to-classical transition of one relevant example: the amplitude
damping channel. We show that, when time goes to infinity, the logical
structure of the system becomes classical.

A quantum operation is a linear and completely positive map from the set of
density operators into itself \cite{Nielsen}. For each time $t$, the quantum
operation $\mathcal{E}_t$ maps the initial state $\hat{\rho}_0$ to the state
at time $t$, i.e.,
\begin{equation}
\mathcal{E}_t(\hat{\rho}_0) = \hat{\rho}(t),
\end{equation}
In terms of the sum representation, we can express $\mathcal{E}_t$ as
follows \cite{Nielsen},
\begin{equation}
\mathcal{E}_t(\hat{\rho}_0) = \sum_{\mu}\hat{E}_{\mu}(t) \hat{\rho}_0 \hat{E}%
^{\dag}_{\mu}(t),
\end{equation}
where $\hat{E}_{\mu}(t)$ are the Kraus operators \cite{KK} associated with
the map $\mathcal{E}_t$.

Now, we define the Heisenberg representation of a quantum operation $%
\mathcal{E}_t$ as an operator $\tilde{\mathcal{E}}_t$ which maps each
observable $\hat {O}$ to another observable $\hat {O}(t) = \tilde{\mathcal{E}%
}_t(\hat {O})$. We interpret $\hat {O}(t)$ as the time evolved observable of
$\hat {O}$ under the quantum operation. The map $\tilde{\mathcal{E}}_t$ must
preserve the mean values of all the observables, i.e.,
\begin{align}
\text{Tr}\left(\hat{\rho}(t) \hat {O}\right) &= \text{Tr}\left(\sum_{\mu}%
\hat{E}_{\mu}(t) \hat{\rho}_0 \hat{E}^{\dag}_{\mu}(t) \hat {O}\right)=
\notag \\
&=\text{Tr}\left(\hat{\rho}_0 \sum_{\mu}\hat{E}^{\dag}_{\mu}(t) \hat {O}
\hat{E}_{\mu}(t)\right)= \text{Tr}\left(\hat{\rho}_0 \hat {O}(t) \right).
\end{align}
Therefore, the map $\tilde{\mathcal{E}}_t$ is given by
\begin{equation}
\tilde{\mathcal{E}}_t(\hat {O})= \hat {O}(t) = \sum_{\mu}\hat{E}%
^{\dag}_{\mu}(t) \hat {O} \hat{E}_{\mu}(t).
\end{equation}
It easy to check that, if $\tilde{\mathcal{E}}_t(\hat {O})$ is a
self-adjoint operator, then $\tilde{\mathcal{E}}_t$ maps the space of
observables into itself.

Once we have defined quantum operations in the Heisenberg picture, we can
study the logical classical limit of the system. Let us illustrate this
process with a simple example: the amplitude damping channel \cite{Nielsen}.
The amplitude damping channel is useful for describing the energy
dissipation of a quantum system due to the effects of an environment. It has
many applications in quantum information processing, because it is
appropriate for modelling the effects of quantum noise. This quantum map can
be used to describe the decay of an excited state of a two-level atom due to
the spontaneous emission of a photon. If the atom is in the ground state
there is no photon emission, and the atom continues in the ground state.
But, if the atom is in the excited state, after an interval of time $\tau$,
there is a probability $p$ that the state has decayed to the ground state
and a photon has been emitted \cite{Nielsen}.

The quantum map which represents the amplitude damping channel can be
expressed in term of two Kraus operators \cite{Nielsen},
\begin{equation}
\mathcal{E}_{\tau}(\hat{\rho}_0) = \hat{E}_0 \hat{\rho}_0 \hat{E}_0^{\dag} +
\hat{E}_1 \hat{\rho}_0 \hat{E}_1^{\dag},
\end{equation}
with the Kraus operators given by
\begin{equation}
\hat{E}_0 = \left(
\begin{array}{ccc}
1 & ~0 &  \\
0 & ~\sqrt{(1-p)} &
\end{array}
\right), ~~~~ \hat{E}_1 = \left(
\begin{array}{ccc}
0 & ~\sqrt{p} &  \\
0 & ~0 &
\end{array}
\right).
\end{equation}
In the Heisenberg picture, we have an associated quantum map $\tilde{%
\mathcal{E}}_{\tau}$ acting on the space of observables, given by $\tilde{%
\mathcal{E}}_{\tau}(\hat{O}) = \hat{E}_0^{\dag} \hat{O} \hat{E}_0 + \hat{E}%
_1^{\dag}\hat{O} \hat{E}_1$. In matrix form, we have the following
expression,
\begin{equation}
\tilde{\mathcal{E}}_{\tau}(\hat{O}) = \left(
\begin{array}{ccc}
O_{00} & ~~~~\sqrt{1-p} \, O_{01} &  \\
\sqrt{1-p}\, O_{10} & ~~~~ p O_{00} + (1-p)O_{11} &
\end{array}
\right).
\end{equation}

If we apply the amplitude damping channel $n$ times, we obtain the quantum
map $\tilde{\mathcal{E}}_{n\tau}(\hat{O})$ given by
\begin{equation}
\tilde{\mathcal{E}}_{n\tau}(\hat{O}) = \left(
\begin{array}{ccc}
O_{00} & ~~~~\sqrt{(1-p)^n} \, O_{01} &  \\
\sqrt{(1-p)^n}\, O_{10} & ~~~~ \sum_{i=0}^{n-1} p (1-p)^i O_{00}+(1-p)^n
O_{11} &
\end{array}
\right).
\end{equation}
This can be reduced to
\begin{equation}
\tilde{\mathcal{E}}_{n\tau}(\hat{O}) = \left(
\begin{array}{ccc}
O_{00} & ~~~~\sqrt{(1-p)^n} \, O_{01} &  \\
\sqrt{(1-p)^n}\, O_{10} & ~~~~(1-p)^n O_{11} + O_{00}\left[1- (1-p)^n \right]
&
\end{array}
\right).
\end{equation}
Considering the limit $n \longrightarrow \infty$, we obtain
\begin{equation}
\tilde{\mathcal{E}}_{\infty}(\hat{O}) = \left(
\begin{array}{ccc}
O_{00} & ~~ 0 &  \\
0 & ~~ O_{00} &
\end{array}
\right).
\end{equation}
Thus, when $t\longrightarrow \infty$, all observables become multiples of
the identity. This means that the whole algebra of observables becomes
trivially commutative, and therefore, the associated lattice becomes Boolean.

The quantum-to-classical transition was extensively studied in the physics
literature from the point of view of the quantum state evolution. However,
from this perspective observables do not evolve on time. In previous papers,
we argued that this kind of descriptions of the classical limit based on the
Schr\"odinger picture is not adequate for explaining the
quantum-to-classical transition of the logical structure of a system.
Instead, the description in terms of the Heisenberg picture allows to
describe how the quantum structure of properties becomes a Boolean.

\section{Unstable systems and rigged Hilbert space}

\label{s:RHS}

In the previous section we have studied the commutation process for a simple case. From a more general perspective, this
phenomenon can appear when the evolution of the system is non-unitary \cite%
{PP,PH,AM}. A natural way of describing this kind of processes has been
largely studied in the literature of resonances and unstable quantum systems
\cite{GP,BO,BO1,AG,CG,RO,ER}. In most of these models, resonances appear
associated with poles of the scattering matrix and give place to decay
times, which can be related with relaxation and decoherence processes \cite%
{deco1,deco2,deco3,deco4,deco5}. The formalism of rigged Hilbert space is a
natural choice for describing these kind of physical processes. In what
follows, we will study the commutation process in the context of this
formalism.

The study of unstable physical systems usually appeals to a master equation or a non-Hermitian Hamiltonian, giving place to a
non-unitary evolution in the Hilbert space \cite{RO,ER, MO,CG2,CGIL}.
In this paper we explore a different approach: we change the Hilbert space for a rigged Hilbert space, in which we obtain an evolution that is suitable for describing unstable systems, and we construct a time evolution operator which is formally
Hermitian although not unitary. This non-unitarity will allow the
evolution from a non-commutative algebra of observables to a commutative one.

Resonance scattering is produced by a Hamiltonian pair $\{H_0,H\}$ with $%
H=H_0+V$. Here, $H_0$ is the so called \textit{free Hamiltonian} and $V$ is
an interaction. If we consider a three dimensional system, usually $H_0=%
\mathbf{p}^2/2m$ and $V$ is given by a spherically symmetric function of the
position $\mathbf{r}$, $V(\mathbf{r})$. For simplicity, we also assume that $%
V(\mathbf{r})$ is short range (vanishes at the infinity faster than the
Coulomb interaction) or even of compact support (it is zero outside a finite
region).

Quasi stationary states are produced when an incoming particle enters into
the interacting region, where the potential $V$ is non-zero, and stays in
this region for a much longer period of time than it would have been if the
interaction were absent.

Quasi stationary states are usually identified with resonances \cite{AFH}.
Resonances are conceptually defined in two ways. We may always assume that
the continuous spectrum of both $H_0$ and $H$ is given by $\mathbb{R}%
^+\equiv [0,\infty)$. For simplicity, we also may assume that both
Hamiltonians do not have bound states (which implies a restriction to the
continuous subspace), singular spectrum or even that the absolutely
continuous spectrum is not degenerate (which in the case of three
dimensional spherically symmetric potentials is equivalent to choose the
subspace with $\ell=0$). Although none of these simplifications is
essential, we will restrict our considerations to Hamiltonians with a
non-singular continuous spectrum.

\medskip

\textit{Definition 1}.- For any pure state $\psi\in\mathcal{H}$ in the
Hilbert space $\mathcal{H}$, let us consider the following pair of complex
functions
\begin{equation}  \label{1}
F_0(z):=\langle \psi|(H_0-z)^{-1}|\psi\rangle, \quad F(z):= \langle
\psi|(H-z)^{-1}|\psi\rangle.
\end{equation}

These functions are meromorphic having the positive semi-axis $\mathbb{R}^+$
as branch cut. Then, if for some $\psi\in\mathcal{H}$, $F(z)$ has a pole at $%
Z_R$ and $F_0(z)$ does not, then we say that the Hamiltonian pair $\{H_0,H\}$
has a resonance at $z_R$ \cite{RS}.

\medskip

\textit{Definition 2}.- Take the $S$ matrix in the momentum representation,
so that $S$ is a function of the modulus, $p:=|\mathbf{p}|$ of the momentum $%
\mathbf{p}$, so that $S\equiv S(p)$. Under some hypothesis related with
causality, $S(p)$ is analytically continuable to the complex plane as a
meromorphic function (that may have additionally branch cuts). The isolated
singularities of this extension are poles (never essential singularities).
Poles on the imaginary axis are always simple. Poles on the positive
imaginary axis represent bound states, poles on the negative imaginary axis
are linked to the existence of antibound or virtual states. Finally, \textit{%
resonances are given by pair of poles on the lower half plane, equidistant
with respect to the imaginary axis. Each of these poles represent a single
resonance. } In principle, there is no restriction with respect to the order
of resonance poles.

\smallskip

It is customary to represent the $S$ matrix in terms of the energy under the
transformation $p=\sqrt{2mE}$. Since the square root is a multiform function
supported on a two sheeted Riemann surface, the same property is shared by
the function $S(E)$ \cite{BOHMBOOK}. On this Riemann surface, resonance
poles appear in complex conjugate pairs and lie on the second sheet.

\medskip

The equivalence of both definitions has not been thoroughly investigated,
although it goes well for some simple models. In addition, there are some
other definitions based on physical notions, which are only equivalent under
additional assumptions \cite{BOHMBOOK,FGR}. We may add that, although the
first definition we give here is widely accepted by mathematicians, the
second one is more popular among physicists. We are using this definition in
the sequel.

In a high number of previous articles, we have discussed the construction of
Gamow vectors in an abstract setting when the potential satisfies the above
mentioned conditions. Let us summarize the main properties of these Gamow
vectors.

\begin{itemize}
\item {Let us consider a resonance defined as a pair of complex conjugate
poles of the analytic continuation of the $S(E)$ matrix in the energy
representation. These poles are located at the points $z_R=E_R-i\Gamma/2$
and $z_R^*=E_R+i\Gamma/2$. Let us assume that these resonance poles are
simple, otherwise unnecessary complications will emerge in the model. The
general theory shows that one may define two vectors, $|\psi^G\rangle$ and $%
|\psi^D\rangle$, related to $z_R^*$ and $z_R$, respectively, with some
properties that we mention in the sequel.}

\item {Both Gamow vectors, $|\psi^G\rangle$ and $|\psi^D\rangle$, are
eigenvectors of the total Hamiltonian $H$ with respective eigenvalues given
by $z_R^*$ and $z_R$, so that
\begin{equation}  \label{2}
H|\psi^G\rangle=z_R^*\,|\psi^G\rangle\,, \qquad H
|\psi^D\rangle=z_R\,|\psi^D\rangle.
\end{equation}
}
These relations define both Gamow vectors.

\item {However, $H$ is a self adjoint Hamiltonian and a self adjoint
Hamiltonian cannot have complex eigenvectors. The situation is saved if we
extend $H$ to the anti-dual space $\Phi^\times$ of a rigged Hilbert space
(RHS in the sequel) $\Phi\subset\mathcal{H}\subset\Phi^\times$. In general,
one may construct two RHS $\Phi_\pm\subset\mathcal{H}\subset \Phi_\pm^\times$
such that $H|\psi^G\rangle=z_R^*\,|\psi^G\rangle$ is valid in $\Phi_-^\times$
and $H |\psi^D\rangle=z_R\,|\psi^D\rangle$ is valid in $\Phi_+^\times$.}

\item {The spaces $\Phi_+$ and $\Phi_-$ are unitarily equivalent to spaces
of complex analytic functions on the upper and lower half planes,
respectively. This construction permits the use of complex analytic function
techniques to obtain our results. In particular, the use of Hardy functions
on a half plane permits a formulation for time asymmetric quantum mechanics
valid for scattering processes. }

\item {We may extend the evolution operator to the antidual spaces $%
\Phi_-^\times$ and $\Phi_+^\times$, so that this operator may be applied to
the Gamow vectors. The result is given by the following pair of relations
\begin{equation}  \label{3}
e^{-itH}|\psi^G\rangle= e^{-itE_R}\,e^{t\Gamma/2}\,|\psi^G\rangle,\quad
e^{-itH}|\psi^D\rangle= e^{-itE_R}\,e^{-t\Gamma/2}\,|\psi^D\rangle.
\end{equation}
Note that $|\psi^G\rangle$ grows and $|\psi^D\rangle$ decays as time
increases in the positive direction. Consequently, $|\psi^G\rangle$ and $%
|\psi^D\rangle$ are named the \textit{growing} and the \textit{decaying}
Gamow vector, respectively. }

\item {When the spaces $\Phi_\pm$ are constructed using Hardy functions, the
first relation in \eqref{3} is valid for $t\le 0$ only. Analogously, the
second relation in \eqref{3} is valid for $t\ge 0$ only. In this formalism,
the unitary group of time evolution splits into two semigroups, one for $%
t\le 0$ and the other for $t\ge 0$. Thus, these RHS supports a semigroup
representation of time evolution.}

\item {Thus, we have two apparently different processes, one for $t\le 0$
and the other for $t\ge 0$. However, the time reversal operator $T$
transforms a process into the other, so that both are essentially
equivalent. In particular,
\begin{equation}  \label{4}
T|\psi^G\rangle=|\psi^D\rangle\,,\qquad T|\psi^D\rangle =|\psi^G\rangle.
\end{equation}%
}

\item {Nevertheless, the basis for time asymmetric quantum mechanics
consists in giving a completely different interpretation to both processes.
Roughly speaking, the RHS $\Phi_-\subset\mathcal{H}\subset \Phi_-^\times$
contains the system observables, while $\Phi_-\subset\mathcal{H}\subset
\Phi_-^\times$ the states. Then, both are different and, thus, time
asymmetry acquires a sense. }
\end{itemize}

In this s
ection we have presented the standard formalism of Rigged Hilbert
spaces. Our aim is to introduce a time evolution operator in order to
describe the evolution of operator algebras in this setting. Thus, we need
to write first the non-Hermitian Hamiltonian in a spectral
decomposition-like expression. This is the subject of the next section.

\section{Generalized time evolution operator in the rigged Hilbert space
formalism}

\label{s:TimeEvolutionInRHS}

In the usual approach to rigged Hilbert space, the dynamical description is
focused on the time evolution of mean values of observables. However, the
expression of a time evolution operator for states (or operators, as seen
from the perspective of the Heisenberg picture) was not present in the
literature. Here we introduce such a time evolution operator. This will
allow us to map non-Abelian algebras into Abelian ones.

\subsection{Non-Hermitian Hamiltonian}

It was shown that any vector $\varphi_+\in\Phi_+$ can be expanded as
\begin{equation}  \label{5}
\varphi_+=\sum_i \alpha_i|\psi_i^D\rangle+|\psi_B\rangle,
\end{equation}
where the sum extends to all resonances, $\alpha_i$ are complex numbers and $%
|\psi_B\rangle$ is the background term. This term is added in order to avoid
a purely exponential decay of normalizable vectors in Hilbert space. Since $%
\Phi_+\subset\Phi_+^\times$, equation \eqref{5} is valid in $\Phi_+^\times$.
Since $\varphi_+$ is normalizable and the Gamow vectors are not, we conclude
that the background term $|\psi_B\rangle$ cannot be normalizable either.

Analogously, for any $\varphi_-\in\Phi_-$, we have the following expansion:
\begin{equation}  \label{6}
\varphi_-=\sum_i\beta_i|\psi_i^G\rangle+|\phi_B\rangle.
\end{equation}

Correspondingly, in \cite{GAD}, we have shown that there are two possible
expansions for the total Hamiltonian $H$ given by the following expressions,
\begin{align}
H&=\sum_i z_{R_i}|\psi_i^D\rangle\langle \psi_i^G|+BGR  \label{7} \\[2ex]
H^\dagger&= \sum_i z^*_{R_i} |\psi_i^G\rangle\langle \psi_i^D|+BGR^*\,.
\label{8}
\end{align}

Both expressions are the formal adjoint of each other. This is why we add
the dagger in the second expansion. They act on different spaces:
\begin{equation}  \label{9}
H\in \mathcal{L}(\Phi_-,\Phi_+^\times),\qquad H^\dagger \in\mathcal{L}%
(\Phi_+,\Phi_-^\times),
\end{equation}
where $\mathcal{L}(\Phi_\pm,\Phi_\mp^\times)$ is the space of continuous
linear mappings from $\Phi_\pm$ to $\Phi_\mp^\times$. Expressions like $BGR$
or $BGR^*$ denote the projection onto the background subspace.

In general, the leading term corresponds to the resonance contribution and
the background in \eqref{7} is usually small. As a matter of fact,
deviations of the exponential law occur for very small and for very large
times and are difficult to be observed. This is why we may omit the
background term for most of observational times. In consequence, a good
approximation for expansions \eqref{7} and \eqref{8} is given if we omit the
background, so that
\begin{equation}  \label{10}
H= \sum_i z_{R_i}|\psi_i^D\rangle\langle \psi_i^G|\,,\qquad H^\dagger=
\sum_i z^*_{R_i} |\psi_i^G\rangle\langle \psi_i^D|.
\end{equation}
Let us insist that the distinction between $H$ and $H^\dagger$ is purely
conventional so that we could have called $H$ or $H^\dagger$ to any of them.

Now, we have the mathematical tools to analyze decoherence produced by
resonances. We show that the above described time evolution gives place to a
commutation process.

\subsection{Time evolution operator}

In order to avoid possible convergence problems, we may assume that the
number of resonances is finite. From the purely mathematical point of view,
this assumption is not fulfilled for most quantum models, but it is still
valid in some useful toy models, such as Friedrichs's one. It is
nevertheless true that resonances with large imaginary terms are not
observable, as the inverse of the imaginary part is related with the mean
life. Also, in the context of a non relativistic theory, large values for
the resonance energy $E_R$ are meaningless. In this way, the approximation
having a finite number of resonances is a reasonable one.

In \cite{GAD}, we have defined a pseudometrics for Gamow vectors. The idea
of using pseudometrics was discussed heuristically in previous articles \cite%
{CGGL,CL}. As mentioned before, let us assume that the number of resonance
poles is finite $\{z_1,z_1^*,z_2,z_2^*,\dots,z_N,z_N^*\}$. Let us consider
the $2N$ dimensional space, $\mathcal{H}^G$, spanned by the Gamow vectors
corresponding to these resonances,
\begin{equation}  \label{11}
\{|\psi_1^D\rangle,|\psi_1^G\rangle, |\psi_2^D\rangle, |\psi_2^G\rangle,
\dots, |\psi_N^D\rangle, |\psi_N^G\rangle \}.
\end{equation}
Notice that $\mathcal{H}^G\subset\Phi_\mp^\times $. We define a
pseudometrics in $\mathcal{H}^G$ by appealing to a matrix:
\begin{equation}  \label{12}
A:= \left(
\begin{array}{ccccccc}
0 & 1 & \dots & \dots & \dots & \dots & \dots \\
1 & 0 & \dots & \dots & \dots & \dots & \dots \\
\dots & \dots & 0 & 1 & \dots & \dots & \dots \\
\dots & \dots & 1 & 0 & \dots & \dots & \dots \\
\dots & \dots & \dots & \dots & \dots & \dots & \dots \\
\dots & \dots & \dots & \dots & \dots & 0 & 1 \\
\dots & \dots & \dots & \dots & \dots & 1 & 0%
\end{array}
\right)\,.
\end{equation}
All entries of $A$ which are not explicitly given are zero. Then, the
pseudoscalar product of two vectors $|\psi\rangle,|\varphi\rangle\in\mathcal{%
H}^G$, $(\psi|\varphi)$, is $(\psi|\varphi)=\langle \psi|A|\varphi\rangle$.
For the basis \eqref{11}, the pseudoscalar products are given by
\begin{eqnarray}  \label{13}
(\psi_i^D|\psi_j^D)=(\psi_i^G|\psi_j^G)=0,\quad
(\psi_i^D|\psi_j^G)=(\psi_i^G|\psi_j^D)=\delta_{ij},
\end{eqnarray}
where $\delta_{ij}$ is the Kronecker delta.

In order to define a sort of time evolution on the space $\mathcal{H}^G$, we
need to use this pseudometrics. First, let us replace the Hamiltonian %
\eqref{10} by
\begin{equation}  \label{14}
H=\sum_{i=1}^N z_i\,|\psi_i^D)( \psi_i^G|.
\end{equation}
More details are discussed in the Appendix. Using the pseudometrics, the
square of $H$ should be
\begin{align}  \label{15}
H^2&=\sum_{i=1}^N z_i\,|\psi_i^D)( \psi_i^G|\sum_{j=1}^N z_j\,|\psi_j^D)(
\psi_j^G|= \sum_{i=1}^N \sum_{j=1}^N z_i\,z_j\, |\psi_i^D)(
\psi_i^G|\psi_j^D)( \psi_j^G|  \notag \\[2ex]
&= \sum_{i=1}^N \sum_{j=1}^N z_i\,z_j\, |\psi_i^D) \delta_{ij} ( \psi_j^G| =
\sum_{i=1}^N z_i^2\, |\psi_i^D) ( \psi_i^G|,
\end{align}
so that
\begin{equation}  \label{16}
H^n= \sum_{i=1}^N z_i^n\, |\psi_i^D) ( \psi_i^G|.
\end{equation}
This allows us to define an expression of the type $e^{-itH}$ as follows
\begin{equation}  \label{17}
U:= e^{-itH}= \sum_{j=1}^N e^{-itz_j}\, |\psi_j^D) ( \psi_j^G|.
\end{equation}
The above expression can be used as time evolution operator on the space of
Gamow vectors. Notice that, in principle, the expression is valid for any
value of $t$. Furthermore, we have $e^{-itH}\in \mathcal{L}%
(\Phi_-,\Phi_+^\times)$.

\section{Commutators}

\label{s:Commutativization}

In this section, we finally deal with the Fort\'in, Holik, Vanni approach
for the decoherence of observables \cite{Fortin2016}. In the first term, we
discuss the simplest case in which a single resonance is present, like in
the basic Friedrichs model. We show that the difference between the
existence of one or more resonances goes beyond the actual complication on
the notation.

We recall that we are studying a resonance scattering process in which the
background term is neglected. Then, our construction is restricted to the
space spanned by the Gamow vectors. In addition, we use the approximation of
having a finite number of resonances (this approximation was motivated in
the previous sections). In this way, the observables under our consideration
are operators acting on the finite dimensional space spanned by the Gamow
vectors.

\subsection{One resonance}

Let $O$ be an observable on the space of Gamow states. Assume that $O$
evolves with time. In this case, we have to define what we understand by
time evolution of an observable. In the case of having just one resonance,
we propose to use the ``complete'' Hermitian Hamiltonian defined as
\begin{equation}  \label{18}
H=z_R|\psi^D)(\psi^G| +z^*_R|\psi^G)(\psi^D|.
\end{equation}
With the aid of the pseudo-metrics \eqref{13}, we obtain the following
expressions
\begin{equation}  \label{19}
H^n= z_R^n |\psi^D)(\psi^G| +(z^*_R)^n|\psi^G)(\psi^D|
\end{equation}
and
\begin{equation}  \label{20}
U(t):=e^{-itH}=e^{-itz_R} |\psi^D)(\psi^G| +e^{-itz_R^*}|\psi^G)(\psi^D|.
\end{equation}

To find the formal inverse of $e^{-itH}$, we need to know an expression for
the identity. Since $\{|\psi^D),|\psi^G)\}$ is a basis in $\mathcal{H}^G$,
let us write
\begin{equation}  \label{21}
I:=|\psi^D)(\psi^G|+|\psi^G)(\psi^D|.
\end{equation}
This is the identity on $\mathcal{H}^G$. Indeed,
\begin{equation}  \label{22}
I|\psi^D)= |\psi^D)(\psi^G|\psi^D)+|\psi^G)(\psi^D|\psi^D) = |\psi^D).
\end{equation}
Analogously, $I|\psi^G)=|\psi^G)$. Then, the linearity of $I$ shows that
this is indeed the identity on $\mathcal{H}^G$. Then, the inverse of $U(t)$
is
\begin{equation}  \label{23}
U(t)^{-1}= e^{itz_R} |\psi^D)(\psi^G| +e^{itz_R^*}|\psi^G)(\psi^D|=U(-t),
\end{equation}
since, using the pseudometric relations \eqref{13}, we have that $%
U(t)=U(-t)=I$.

Next, the time evolution of the operator, which at time $t=0$ is $O$ is now
\begin{equation}  \label{24}
O(t)=U(t)O U(-t).
\end{equation}

We have just one resonance with resonance pole at $z_R=E_R-i\Gamma/2$, so
that the above equation reads
\begin{align}  \label{25}
O(t)&= \left[ e^{-itz_R}|\psi^D)(\psi^G|+e^{-itz_R^*}|\psi^G)(\psi^D|\right]
O \left[e^{itz_R}|\psi^D)(\psi^G|+ e^{itz_R^*} |\psi^G)(\psi^D|\right] =
\notag \\
&= |\psi^D)(\psi^G|O|\psi^D)(\psi^G|+
e^{-t\Gamma}|\psi^D)(\psi^G|O|\psi^G)(\psi^D| +  \notag \\
&~~~+e^{t\Gamma}
|\psi^G)(\psi^D|O|\psi^D)(\psi^G|+|\psi^G)(\psi^D|O|\psi^G)(\psi^D|,
\end{align}
where the ``averages'' $(\psi^G|O|\psi^G)$, etc are in principle well
defined, since we work on a finite dimensional space. In fact, the dimension
is 2 in this case.

This provides undesirable terms in $e^{t\Gamma}$ with $t>0$, so that we have
to give up the condition $U(t)U(-t)=I$. A second choice for the time
evolution of the observables will include the following ingredients:
\begin{equation}  \label{26}
U(t)= e^{-itz_R}|\psi^D)(\psi^G|+e^{itz_R^*}|\psi^G)(\psi^D|.
\end{equation}
The point is that this operator is formally Hermitic. Clearly, its square is
not the identity, instead
\begin{equation}  \label{27}
U(t)U^\dagger(t)=U^2(t)=e^{-t\Gamma} I.
\end{equation}

This choice provides more desirable results. The standard definition of the
time evolution for an observable states that the value of the observable $O$
after a time $t$ is given by $O(t):= U^\dagger(t)OU(t)$. In our case, $%
U^\dagger(t)=U(t)$. Note that the commutator
\begin{align}  \label{28}
[O_1(t),O_2(t)]&=U(t)O_1U(t)U(t)O_2U(t)-U(t)O_2U(t)U(t)O_1U(t) =  \notag \\
&= e^{-t\Gamma}[ U(t)O_1O_2U(t)-U(t)O_2O_1U(t) ]=  \notag \\
&= e^{-t\Gamma} U(t)[O_1,O_2]U(t).
\end{align}

Using the above machinery, we may calculate $U(t)[O_1,O_2]U(t)$. It is a sum
of four terms:
\begin{align}
&e^{-2iE_R}\,e^{-t\Gamma}\, |\psi^D)(\psi^G|[O_1,O_2]|\psi^D)(\psi^G|,
\notag \\
&e^{2iE_R}\, e^{-t\Gamma}\, |\psi^G)(\psi^D|[O_1,O_2]|\psi^G)(\psi^D|,
\notag \\
&e^{-t\Gamma}\, |\psi^D)(\psi^G|[O_1,O_2]|\psi^D)(\psi^G|,  \notag \\
&e^{-t\Gamma}\, |\psi^G)(\psi^D|[O_1,O_2]|\psi^G)(\psi^D|.  \notag
\end{align}
The terms of the form $(\psi^G|[O_1,O_2]|\psi^D)$ and $(\psi^D|[O_1,O_2]|%
\psi^G)$ should be well defined as they are ``averages'' of linear operators
on finite dimensional spaces. We realize that \eqref{28} is of the form:
\begin{equation}  \label{33}
[O_1(t),O_2(t)]= e^{-2t\Gamma}\,\{\alpha(t) |\psi^D)(\psi^G|+ \beta(t)
|\psi^G)(\psi^D|\},
\end{equation}
where $\alpha(t)$ and $\beta(t)$ are constants for which the dependence on $t
$ is just a phase of the form $e^{\pm 2itE_R}$ with $E_R$ real. We have
obtained the result given in \cite{Fortin2016}.

\subsection{More than one resonance}

Here, the procedure is the same, although calculations are more cumbersome.
In general, one should obtain
\begin{eqnarray}  \label{34}
[O_1(t),O_2(t)]= \sum_{j=1}^N e^{-2t\Gamma_j}\,\{\alpha_j(t)
|\psi_j^D)(\psi_j^G|+ \beta_j(t) |\psi_j^G)(\psi_j^D|\},
\end{eqnarray}
where the resonance poles are located at the points $z_j=E_j-i\Gamma_j/2$, $%
j=1,2,\dots,N$.

\subsection{Compatibility with TAQM}

It is important to remark that the formalism presented in this work can be
applied to a wide family of physical models of interest. As an example, we
show in this section the compatibility with the TAQM formalism \cite{BO2,BGK}%
, which finds applications in scattering resonances and more recently in
classical and quantum optics \cite{SO,MAR}.

Let us go to \eqref{26} and observe the coefficients $e^{-itz_R}$ for $%
|\psi^G)(\psi^D|$ and $e^{itz_R^*}$ for $|\psi^D)(\psi^G|$. Also note that,
in the standard formulation of TAQM using RHS of Hardy functions on a half
plane, we have the following evolution rules:
\begin{equation}  \label{35}
e^{-itH}|\psi^D)=e^{-itz_R}|\psi^D)\,,\qquad t\ge 0,
\end{equation}
and
\begin{equation}  \label{36}
e^{-itH}|\psi^G)=e^{-itz_R^*}|\psi^G)=e^{i(-t)z_R^*}|\psi^G)\,,\qquad t\le 0.
\end{equation}

If we want to have a forward time evolution, we use the conversion $%
-t\longmapsto t$ in \eqref{36}. Thus, for $t\ge 0$, the expression $%
e^{itz_R}|\psi^G)$ has full sense. Contrary, for $t\le 0$ time evolution for
these Gamow vectors does not exist in the context of TAQM. Therefore,
equations (\ref{35},\ref{36}) are valid for $t\ge 0$ only from this point of
view.

\section{Conclusions}

\label{s:Conclusions}

In previous works \cite{Fortin2014,Fortin2016,Losada2017}, we have studied
the quantum-to-classical transition from the point of view of the algebra of
observables of the system. If a quantum system undergoes a physical process
such that its behavior becomes classical, then its algebra of observables
should undergo a process from a non-Abelian algebra to an Abelian one.

In this paper, we continue this approach. We introduce a dynamical evolution
operator for dealing with unstable physical process (such as scattering
resonances, photon emission, decoherence and particle decay). In order to do
this, we use the formalism of rigged Hilbert space and we represent the
time evolution of quantum observables in the Heisenberg picture, in such a
way that time evolution is non-unitary. This allows us to describe observables
that are initially non-commutative, but become commutative after time
evolution. Therefore, we show that the quantum-to-classical transition
based in dynamical algebras, occurs in a rich family of models of unstable
systems.

\section*{Appendix}

The replacement of \eqref{10} by \eqref{14} is the
change of $H$ given by \eqref{10} by $BHB$, where $B$ is a square root of $A$%
. As a matter of fact, this means the use of a new operator $H$ of the form:
\begin{equation}  \label{37}
H=\sum_{i=1}^N z_i\,B |\psi_j^D\rangle \langle \psi_j^G| B,
\end{equation}
where $B$ is not uniquely defined. We may choose the following definition
for $B$: replace the $2\times 2$ dimensional nonvanishing boxes in $A$ by
\begin{equation}  \label{38}
(-i)^{1/2}\left(
\begin{array}{cc}
i\sqrt2/2 & \sqrt2/2 \\[2ex]
\sqrt2/2 & i\sqrt2/2%
\end{array}
\right)\,.
\end{equation}
Note that $B|\psi_i^D\rangle=|\psi_i^D)$ and $\langle\psi_i^G|B=(\psi_i^G|$.
The square of $H$ as in \eqref{37} is given by
\begin{align}  \label{39}
H^2&=\sum_{i=1}^N\sum_{j=1}^N z_iz_j B|\psi_i^D\rangle\langle \psi_i^G|BB
|\psi_j^D\rangle\langle \psi_j^G|B =  \notag \\
&= \sum_{i=1}^N\sum_{j=1}^N z_iz_j B|\psi_i^D\rangle\langle \psi_i^G|A
|\psi_j^D\rangle\langle \psi_j^G|B =  \notag \\
&=\sum_{i=1}^N\sum_{j=1}^N z_iz_j B|\psi_i^D\rangle\,\delta_{ij}\,\langle
\psi_j^G|B = B \left[\sum_{i=1}^N z_i^2 |\psi_i^D\rangle\langle \psi_j^G|%
\right] B.
\end{align}
Thus,
\begin{equation}  \label{40}
e^{-itH}= B\left[ \sum_{i=j}^N e^{-itz_j} |\psi_j^D\rangle\langle \psi_j^G| %
\right] B.
\end{equation}

In relation with the formal adjoint $H^\dagger$, it seems convenient to use
another square root of $A$, that we call $C$. Formally, $C$ is the
adjoint of $B$, $C:=B^\dagger$. It is a square root of $A$ since $B^\dagger
B^\dagger=(BB)^\dagger=A^\dagger=A$. Then, the new $H^\dagger$ would be
\begin{equation}  \label{41}
H^\dagger =C\left[ \sum_{i=1}^N z_i^* |\psi_i^G\rangle\langle \psi_i^D| %
\right] C=\sum_{i=1}^N z_i^* |\psi_i^G)(\psi_i^D|.
\end{equation}
Then, $|\psi_i^G)=C|\psi_i^G\rangle$ and $(\psi_i^D|C=\langle \psi_i^D|$.

This choice has an interest by its own. In fact, note that the following
expression is formally Hermitian:
\begin{equation}  \label{42}
H=\sum_{i=1}^N z_i|\psi_i^D\rangle\langle\psi_i^G|+\sum_{j=1}^N z_j^*
|\psi_j^G\rangle\langle \psi_j^D|.
\end{equation}
Then, the formal hermiticity of
\begin{eqnarray}  \label{43}
H=\sum_{i=1}^N z_i|\psi_i^D)(\psi_i^G|+\sum_{j=1}^N z_j^* |\psi_j^G)(
\psi_j^D|
\end{eqnarray}
requires that \eqref{43} be equal to
\begin{equation}  \label{44}
H= B\left[ \sum_{i=1}^N z_i|\psi_i^D\rangle\langle\psi_i^G|\right] B+C\left[%
\sum_{j=1}^N z_j^* |\psi_j^G\rangle\langle \psi_j^D|\right] C.
\end{equation}

\section*{Acknowledgements}

M. Gadella wishes to acknowledge financial support to the Ministerio de
Econom\'ia y Competitividad of Spain (Project MTM2014-57129-C2-1-P with
EU-FEDER support) and the Junta de Castilla y Le\'on (Project VA057U16).
Also to Prof. Olimpia Lombardi and her group for stimulating discussions as
well as for their warm hospitality in Buenos Aires.

S. Fortin, F. Holik and M. Losada wish to acknowledge the financial support of the Universidad de Buenos Aires, the grant 57919 from the John Templeton Foundation and the grant PICT-2014-2812 from the Consejo Nacional de Investigaciones Cient\'ificas y T\'ecnicas of Argentina.

~

\end{document}